\documentclass[11pt]{article}
\usepackage{BSMvietnam,epsfig}

\def\be{\begin{equation}}
\def\ee{\end{equation}}
\def\bea{\begin{eqnarray}}
\def\eea{\end{eqnarray}}

\begin{document}
\vspace*{4cm}
\title{New SUSY Thought}

\author{Hyung Do Kim}

\address{Department of Physics and Astronomy and Center for Theoretical Physics,\\
Seoul National University, Seoul, 151-747, Korea}

\maketitle\abstracts{
New SUSY thought is presented.
Maximal stop mixing needed for 125 GeV Higgs is linked to the tachyonic stop at the UV boundary. Large $\mu$ does not mean the severe fine tuning if Higgs comes out as a pseudo-Goldstone boson. The small mass of the pseudo-Goldstone Higgs is overcome with extra vector-like fermions needed to explain the Higgs to di-photon rates.
}

\section{Introduction}

The Large Hadron Collider (LHC) experiments run successfully and provide excellent data.
Weak scale supersymmetry confronts two challenges from the LHC experiments.
Firstly, no discovery of superparticles yet at the LHC is a puzzle 
for those who predicted superparticle mass at the weak scale or below TeV.
Now the mass bounds for gluino and squark went up over 1 TeV.
Secondly, the discovery of the Standard Model Higgs-like new boson with its mass
at around 125 GeV \cite{:2012gk} \cite{:2012gu} is quite difficult to be explained in the weak scale supersymmetry.

Nevertheless, at least two facts are consistent with each other. Lighter Higgs mass requires 
lighter superparticles and no observation of superparticles could have ruled out the possibility of weak scale supersymmetry.

With the limits from the current direct search and the observed Higgs mass,
a few nonstandard supersymmetric ideas are introduced with a personal taste.
Natural supersymmetry, Higgs as a pseudo-Goldstone boson and the Higgs to di-photon rate are mainly discussed.

\section{Natural supersymmetry}

Natural supersymmetry \cite{Asano:2010ut} \cite{Papucci:2011wy} \cite{Hall:2011aa}, or more precisely, the least unnatural supersymmetry is motivated by the natural (least unnatural) electroweak symmetry breaking.
It can be realised with light stop while avoiding the direct search bound applied to the first two generations which are not directly related to the electroweak symmetry breaking.
The largest finite threshold correction after integrating out stops is possible when two stops have the maximal mixing with $|A_t/m_{\tilde{t}}| \sim \sqrt{6}$ where $A_t$ is the soft tri-linear coupling for Higgs and stops and $m_{\tilde{t}}$ is the soft stop mass \cite{Carena:2000dp}.

There are problems with the maximal stop mixing if the UV (ultra-violet) completion of the theory is considered.
First of all, it can not be obtained in gauge mediation.
Minimal gauge mediation does not generate $A_t$ term at the messenger scale (at one loop).
Therefore, the weak scale $A_t$ term is generated from the gluino loop
and $A_t/m_{\tilde{t}}$ can not be bigger than 1 as the same gluino loop increases the stop mass.

Secondly, it is also difficult to realise the maximal stop mixing within the conventional SUSY breaking models in which the mediation scale is at around the Grand Unified Theory (GUT) scale, $M_{\rm GUT} = 2 \times 10^{16}$ GeV or at the Planck scale, $M_{\rm Pl} = 2\times 10^{18}$ GeV.
The electroweak scale soft parameters can be written in terms of the GUT scale parameters as follows  \cite{Dermisek:2006ey}. (Here we only consider gluino mass, stop mass and $A_t$ as they provide the leading contribution in the running unless extremely different hierarchy is imposed between the parameters at the GUT scale.)
\begin{eqnarray}
m_{\tilde{t}}^2 (M_Z) & \simeq & 5.0 M_3^2 + 0.6 m_{\tilde{t}}^2 + 0.2 A_t M_3, \nonumber \\
M_3 (M_Z) & \simeq & 3 M_3, \nonumber \\
A_t (M_Z) & \simeq & -2.3 M_3 + 0.2 A_t, \nonumber
\end{eqnarray}
where $M_3$ is the gluino mass, $m_{\tilde{t}}$ is the stop mass and $A_t$ is the soft tri-linear coupling between Higgs and stops.
The boundary $A_t$ term is exponentially suppressed at low energy and the weak scale $A_t$ term comes from the running.
To overcome it, $A_t$ at the GUT scale should be at least 5 or 10 times larger than the stop mass.

There are two ways out.
Firstly, the tachyonic stop boundary condition at high energy scale can realise the maximal stop mixing \cite{Dermisek:2006ey} \cite{Dermisek:2006qj}. The stop mass squared beomes positive at low energy even if we start from negative value at high energy. $A_t$ can be of order of the gluino mass at the weak scale.
If the stop mass squared changes sign close to the weak scale, the stop mass can be very small at the weak scale compared to the gluino mass and $A_t/m_{\tilde{t}}$ can be larger than 1.
In this scenario, it is essential that the stop is tachyonic at the UV boundary.
Secondly, the messenger scale is not high and $A_t$ term is generated at the messenger scale.
It is possible in Yukawa assisted gauge mediation. 

\begin{figure}[htb]
  \begin{center}
   \includegraphics[width=0.65\textwidth]{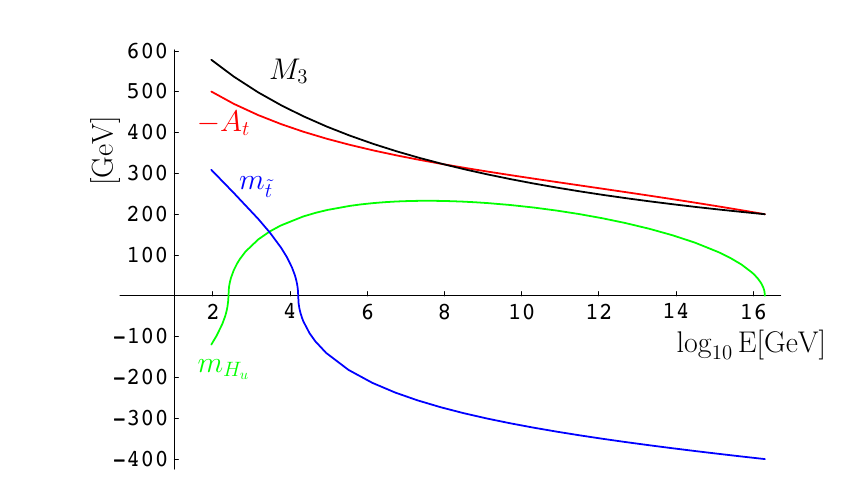}
  \end{center}
 \caption{RG running of soft SUSY breaking parameters for $\tan \beta =10$ 
 and GUT scale boundary conditions : $-A_t=M_3=200$ GeV, $m_{\tilde{t}}^2=-(400{\ \rm GeV})^2$ and $m_{H_u}^2=0 {\ \rm GeV^2}$. $m = m^2/\sqrt{|m^2|}$ is used for soft scalar mass.}
\label{fig:RGstop}
\end{figure}

The Fig. \ref{fig:RGstop} is taken from \cite{Dermisek:2006ey}.
The current bound on the gluino mass is about 1 TeV and the y-axis scale should be multiplied by factor 2 to be consistent with the current limit. For the discussion given here, only the relative size matters. It is clear that the ratio of $|A_t|/m_{\tilde{t}}$ can be very large at just below the transition scale (10 TeV in the plot). Therefore, the maximal stop mixing can be easily realised if the stop has a tachyonic boundary condition. 

\begin{figure}[htb]
  \begin{center}
   \includegraphics[width=0.65\textwidth]{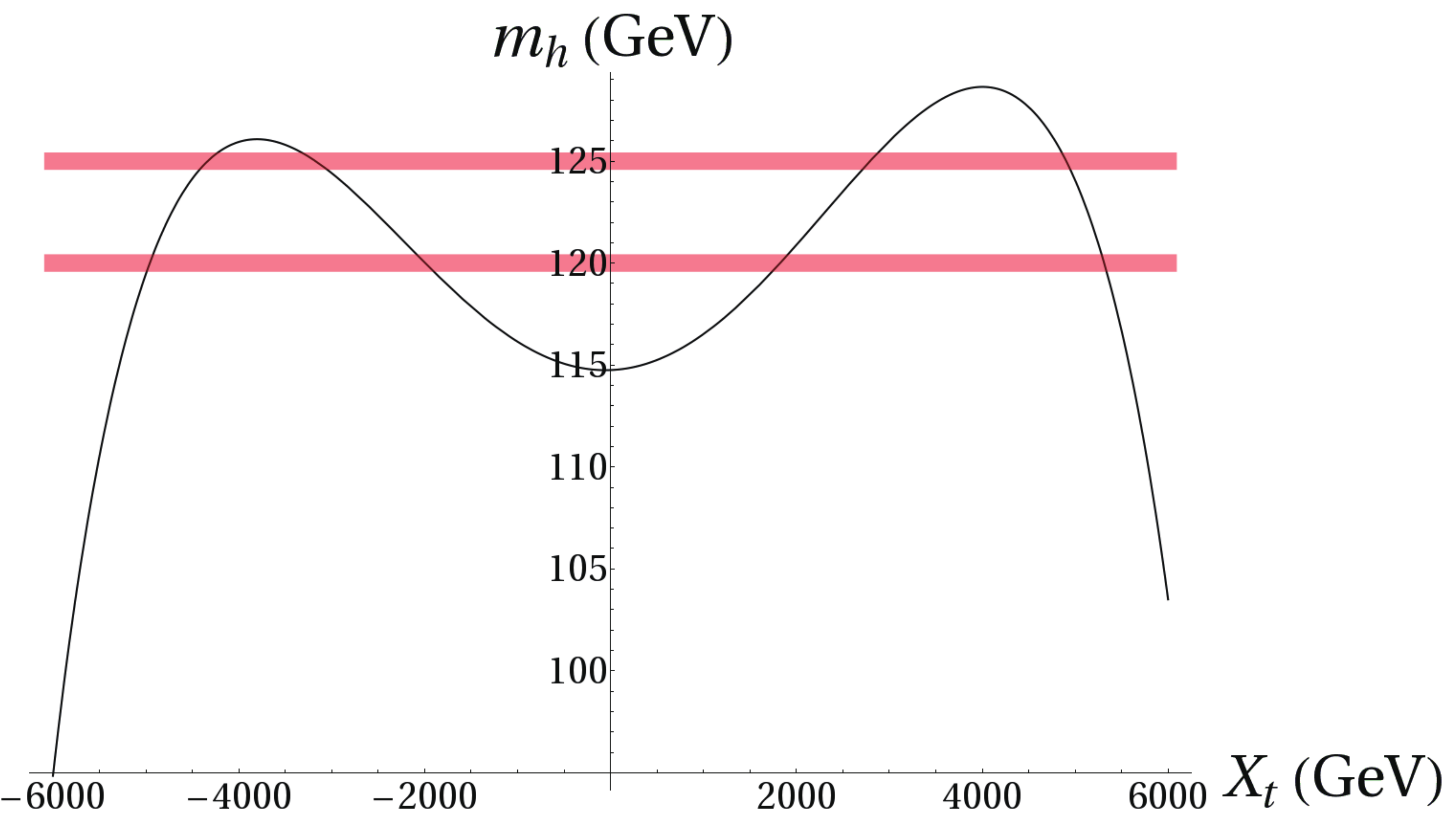}
  \end{center}
 \caption{Higgs mass as a function of $X_t=A_t-\mu/\tan \beta$ with $\tan \beta=10$, $m_{\tilde{t}}=2$ TeV}
\label{fig:maximalmixing}
\end{figure}

The Fig. \ref{fig:maximalmixing} is taken from \cite{Kim:2012vz}. Compared with the case $|X_t/m_{\tilde{t}}| \sim 1$,
the maximal stop mixing can increase the Higgs mass by 5 GeV for the same stop mass. Though 5 GeV is a small addition to 125 GeV, this small addition is crucial here since otherwise the stop should be at least 5 TeV to raise the Higgs mass solely from the logarithmic running of the quartic coupling.

\begin{figure}[htb]
  \begin{center}
   \includegraphics[width=0.65\textwidth]{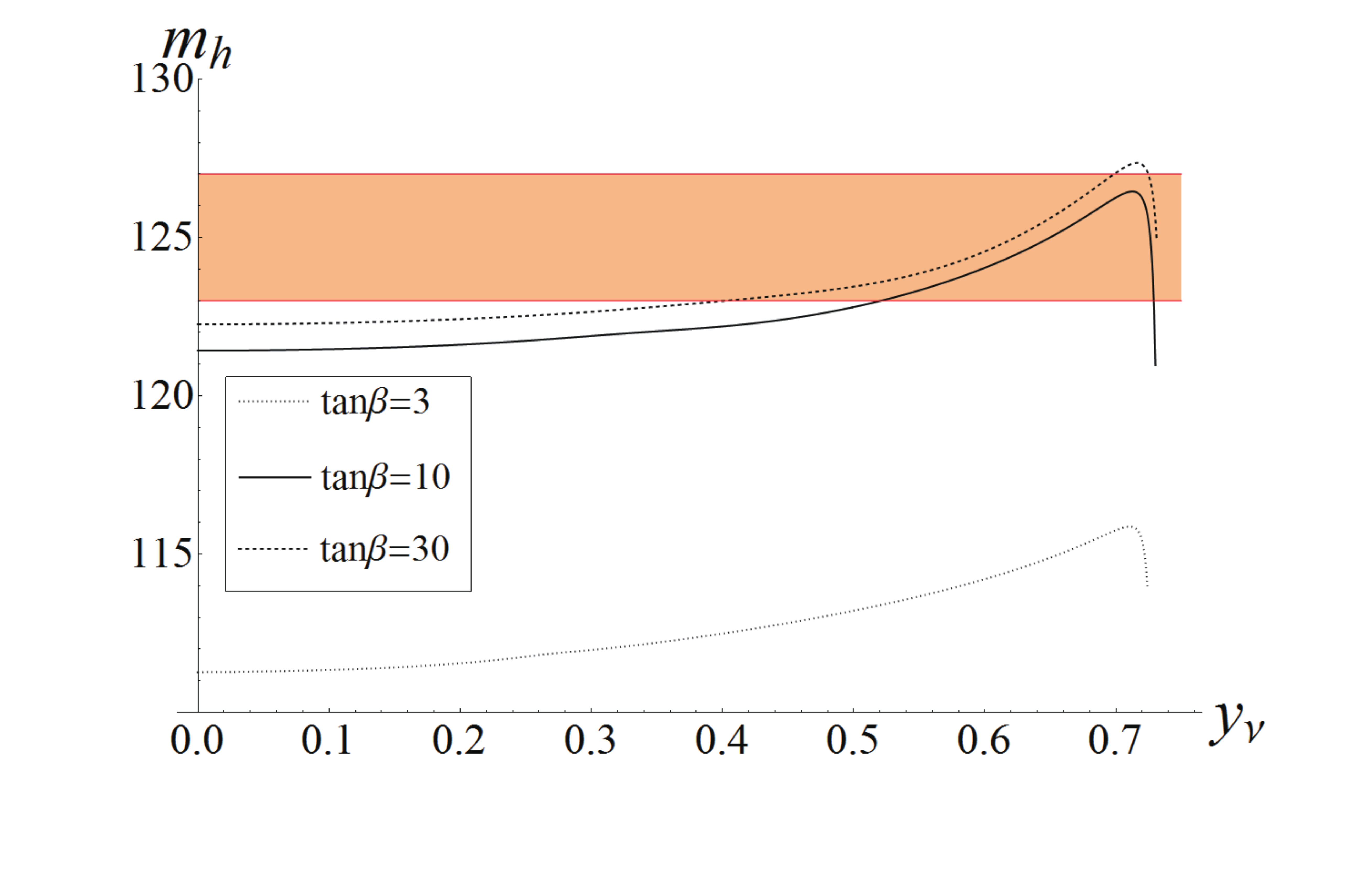}
  \end{center}
 \caption{Higgs mass as a function of $y_{\nu}$ for  $B_N = F/M = 5  \times 10^{5}$ GeV, $\rho = 0.1$. Higgs mass can be achieved with the help of Yukawa mediation for large $\tan \beta$ region. At $y_{\nu} \sim 0.7$, stop mass is approximately $1$ TeV.}
\label{fig:mh}
\end{figure}

The Fig. \ref{fig:mh} is also taken from \cite{Kim:2012vz} in which the 3 right-handed neutrinos are considered as the additional messengers of supersymmetry breaking. In the $S_4$ flavour model discussed in \cite{Kim:2012vz}, the maximal stop mixing is realised for neutrino Dirac Yukawa coupling $y_\nu \sim 0.7$. It is possible since the stop becomes tachyonic at the See-Saw scale.
It can be made to be consistent with the $\mu \to e \gamma$ bounds by special construction but it is very difficult to accommodate the muon $g-2$ in minimal neutrino assisted gauge mediation.

\section{$\mu/B\mu$ problem}

In the weak scale supersymmetric theory, $\mu$ term is the only supersymmetric mass term that is allowed in the superpotential. The correct electroweak symmetry breaking requires $\mu$ to be comparable to the SUSY breaking parameters or smaller than those. The weak scale arises as a cancellation between $\mu^2$ and other soft parameters. If $\mu^2$ is much larger than other soft SUSY breaking parameters, there would be no electroweak symmetry breaking allowed as the Higgs potential would be stable at the origin with the help of large $\mu^2$. Direct search for chargino puts a limit on the lowest possible value of $\mu$ to be larger than 100 GeV. Therefore, in weak scale supersymmetric theory, $\mu$ term should be of similar size of the soft SUSY breaking parameters.
$\mu$ problem is why the supersymmetric mass term $\mu$ should be comparable to other soft supersymmetry breaking parameters \cite{Kim:1983dt}. The connection can be explained by Giudice-Masiero mechanism \cite{Giudice:1988yz}.
However, it works only when the scale of the messengers is comparable to the Planck scale.
One way out is to generate the $\mu$ term from the non-renormalizable interactions.
As a result we can connect the strong CP problem and the $\mu$ problem \cite{Kim:1983dt} \cite{Choi:2011rs}.
The other possibility is to couple Higgs to the messengers of supersymmetry breaking.
It is easy to generate $\mu$ at one loop like other soft supersymmetry breaking parameters,
but at the same time $B\mu$ is generated at one loop unless special model building care is taken \cite{Dvali:1996cu} \cite{Giudice:2007ca}.
One loop generated $\mu$ and two loop generated $B\mu$ are realised in \cite{Dvali:1996cu} using supercovariant derivative term and in  \cite{Giudice:2007ca} using a single field dominance
to isolate holomorphic term from anti-holomorphic term.
The extension of the MSSM to the NMSSM \cite{Delgado:2007rz} can possibly provide a solution to the $\mu$ problem
as there is an extra singlet which can take a vacuum expectation value (VEV) and can generate $\mu$ without other phenomenological problems.
In summary successful gauge mediation model should solve $\mu/B\mu$ problem.
The solution of $\mu/B\mu$ in general makes the model very complicated.
Therefore, to build a successful gauge mediation model, it would be better to start from the solution of $\mu/B\mu$.

\section{Higgs as a pseudo-Goldstone boson}

In most of the fine tuning analysis, the fine tuning is computed as a ratio of $M_Z^2$ and $m_H^2$.
More precisely it should be the ratio of the physical Higgs mass, $m_h^2$ and the soft scalar mass $m_H^2$. 
\begin{eqnarray}
\frac{m_h^2}{2} & \simeq & -\mu^2 - m_{H_u}^2, \nonumber
\end{eqnarray}
for $\tan \beta \gg 1$.

The fine tuning \cite{Barbieri:1987fn} is defined as $\Delta_{\mu} = \frac{2\mu^2}{m_h^2}$.
The underlying reason is that the generation of $\mu$ involves the additional coupling which is nothing to do with the mechanism of supersymmetry breaking. $\Delta_{\mu}$ can be regarded as the least amount of fine tuning needed.

The loophole in the argument arises if the light Higgs is a pseudo-Goldstone boson.
In this case the light Higgs potential is predicted to be flat independently of $\mu$ 
at the scale of global symmetry breaking.
The light Higgs mass is generated from the loop correction proportional to top Yukawa and gauge couplings and can be independent of $\mu$ in this case.

The pseudo-Goldstone boson Higgs has been extensively studied in the context of supersymmetry, for instance, look at \cite{Birkedal:2004xi} \cite{Bae:2012ir}. In \cite{Bae:2012ir}, the idea of Dvail, Giudice and Pomarol \cite{Dvali:1996cu} has been used to realise the setup in which Higgs appears as a pseudo-Goldstone boson after the global symmetry is broken down to its subgroup. 
To realise the accidental global symmetry at the scale at around 10 TeV is a difficult task 
but we postpone the discussion to the future work and focus on the consequence afterwards.
$\mu$ term is generated when the global symmetry is broken. In this model $\tan \beta=1$ is predicted and the Higgs potential vanishes at the global symmetry breaking scale.
The pseudo-Goldstone boson predicts light Higgs much lighter than 120 GeV since the tree level contribution to the quartic coupling vanishes.

\section{Higgs to di-photon rate}

On the other hand the slight excess of the Higgs to di-photon rate in ATLAS \cite{:2012gk}
and CMS \cite{:2012gu} needs extra charged particles which are nothing to do with the hierarchy problem. Indeed in most of the natural theory in which new coloured particles cancel the top loop in the Higgs mass, the Higgs to di-photon rate is expected to be suppressed rather than enhanced \cite{Low:2009di}. The fermionic top partner which cancels the quadratic divergence in the Higgs mass also suppresses the gluon fusion diagram to the Higgs. As a result less number of Higgs events are expected. The scalar top partner can add up the gluon fusion if there is no mixing but the gluon fusion to the Higgs is suppressed if maximal stop mixing is taken into account \cite{Low:2009di}.
The change in Higgs to di-photon diagram is small as the leading diagram is the W boson loop.
As a result, in motivated theory for the hierarchy problem, the Higgs to di-photon rate which comes out as the product of gluon fusion and Higgs to di-photon is predicted to be suppressed.

The enhancement of the Higgs to di-photon rate needs an ad hoc introduction of extra charged particles \cite{Carena:2012xa}. Furthermore, these extra charged particles should have order one Yukawa couplings with the Higgs to affect the Higgs to di-photon rate significantly.

Once we accept the vector-like charged fermions to explain the di-photon enhancement, the same Yukawa can contribute to the RG running of the Higgs quartic coupling. Therefore, with extra vector-like fermions, we can accommodate the pseudo-Goldstone boson Higgs in supersymmetry.
Furthermore, the stop mixing parameter is simply written as $X_t = A_t - \mu/\tan \beta = A_t -\mu$ for $\tan \beta=1$.
Now we can realise the maximal stop mixing with large $\mu$ even in the absence of $A_t$
as $X_t=-\mu$ \cite{Bae:2012ir}.

\begin{figure}[htb]
  \begin{center}
   \includegraphics[width=0.65\textwidth]{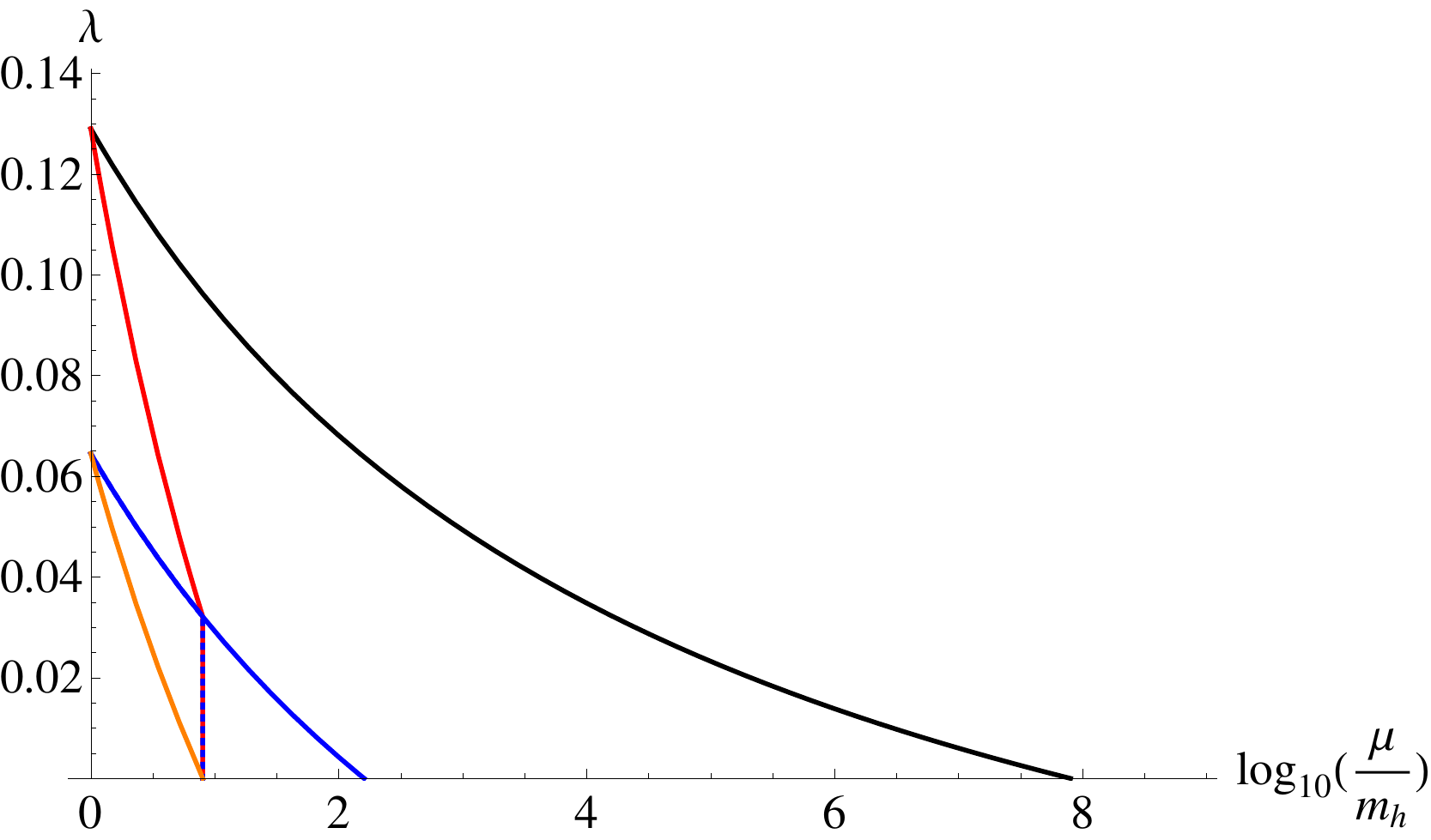}
  \end{center}
 \caption{Running of Higgs quartic coupling in various scenarios. 
 The black line corresponds to the Standard Model. In the MSSM, $\tan \beta=1$ corresponds to vanishing quartic at tree level.
 The blue line shows the slope given by top Yukawa. The orange line shows the slope with vector-like fermions. The red line represents the scenario with $\tan \beta =1$ with vector-like fermions and maximal stop mixing.}
\label{fig:quartic}
\end{figure}

The Fig. \ref{fig:quartic} is taken from \cite{Bae:2012ir}. The red line shows that $\tan \beta=1$ and stop mass at around 1 TeV can explain the observed 125 GeV Higgs mass with the help of vector-like charged fermions. It is also possible to increase the Higgs mass using the vector-like fermions while suppressing the contribution to the Higgs to di-photon rate.

To separate the scale of the physical Higgs mass from $\mu$ in the framework of pseudo-Goldstone boson is an interesting possibility to look for in the future.

\section{Conclusion}

The observation of the SM Higgs-like new boson with mass at around 125 GeV challenges the weak scale supersymmetry. The most popular maximal mixing scenario is hard to make the UV completion. Any mediation of SUSY breaking which can realise the tachyonic stop at the UV boundary can realise the maximal stop mixing. The recent construction of 'neutrino assisted gauge mediation' \cite{Kim:2012vz} does the job.

Large $\mu$ is regarded as a sign of severe fine tuning. However, it can be overcome if the Higgs comes as a pseudo-Goldstone boson when $\mu$ is generated after the global symmetry breaking.
The predicted mass of the Higgs boson can be too light to be consistent with the observed one.
Nevertheless, extra vector-like charged fermions can help raise the Higgs quartic coupling just like the top. Extra vector-like charged fermions are motivated by the enhancement of the Higgs to di-photon rate.

The current Higgs signal is puzzling for the natural theories. Natural theories generally predict the suppressed events due to the cancellation in the gluon fusion loop. If the di-photon enhancement persists, it would be a clear indication of new charged particles at around a few 100 GeV. These particles are not motivated by the gauge hierarchy problem.

\section*{Acknowledgments}
This work is supported by the NRF of Korea No. 2011-0017051.
HDK thanks to the organisers of the BSM 2012 conference in Vietnam.



\section*{References}
\begin{thebibliography}{99}

\bibitem{:2012gk} 
  G.~Aad {\it et al.}  [ATLAS Collaboration],
  Phys.\ Lett.\ B {\bf 716}, 1 (2012)
  [arXiv:1207.7214 [hep-ex]].
  
\bibitem{:2012gu} 
  S.~Chatrchyan {\it et al.}  [CMS Collaboration],
  Phys.\ Lett.\ B {\bf 716}, 30 (2012)
  [arXiv:1207.7235 [hep-ex]].

\bibitem{Asano:2010ut} 
  M.~Asano, H.~D.~Kim, R.~Kitano and Y.~Shimizu,
  JHEP {\bf 1012}, 019 (2010)
  [arXiv:1010.0692 [hep-ph]].

\bibitem{Papucci:2011wy} 
  M.~Papucci, J.~T.~Ruderman and A.~Weiler,
  JHEP {\bf 1209}, 035 (2012)
  [arXiv:1110.6926 [hep-ph]].

\bibitem{Hall:2011aa} 
  L.~J.~Hall, D.~Pinner and J.~T.~Ruderman,
  JHEP {\bf 1204}, 131 (2012)
  [arXiv:1112.2703 [hep-ph]].
  
\bibitem{Carena:2000dp} 
  M.~S.~Carena, H.~E.~Haber, S.~Heinemeyer, W.~Hollik, C.~E.~M.~Wagner and G.~Weiglein,
  Nucl.\ Phys.\ B {\bf 580}, 29 (2000)
  [hep-ph/0001002], and references therein.

\bibitem{Dermisek:2006ey} 
  R.~Dermisek and H.~D.~Kim,
  Phys.\ Rev.\ Lett.\  {\bf 96}, 211803 (2006)
  [hep-ph/0601036].
  
\bibitem{Dermisek:2006qj} 
  R.~Dermisek, H.~D.~Kim and I.~-W.~Kim,
  JHEP {\bf 0610}, 001 (2006)
  [hep-ph/0607169].
  
\bibitem{Kim:2012vz} 
  H.~D.~Kim, D.~Y.~Mo and M.~-S.~Seo,
  arXiv:1211.6479 [hep-ph].
  
\bibitem{Giudice:1988yz} 
  G.~F.~Giudice and A.~Masiero,
  Phys.\ Lett.\ B {\bf 206}, 480 (1988).

\bibitem{Kim:1983dt} 
  J.~E.~Kim and H.~P.~Nilles,
  Phys.\ Lett.\ B {\bf 138}, 150 (1984).

\bibitem{Choi:2011rs} 
  K.~Choi, E.~J.~Chun, H.~D.~Kim, W.~I.~Park and C.~S.~Shin,
  Phys.\ Rev.\ D {\bf 83}, 123503 (2011)
  [arXiv:1102.2900 [hep-ph]].
  
\bibitem{Dvali:1996cu} 
  G.~R.~Dvali, G.~F.~Giudice and A.~Pomarol,
  Nucl.\ Phys.\ B {\bf 478}, 31 (1996)
  [hep-ph/9603238].

\bibitem{Giudice:2007ca} 
  G.~F.~Giudice, H.~D.~Kim and R.~Rattazzi,
  Phys.\ Lett.\ B {\bf 660}, 545 (2008)
  [arXiv:0711.4448 [hep-ph]].
  
\bibitem{Delgado:2007rz} 
  A.~Delgado, G.~F.~Giudice and P.~Slavich,
  Phys.\ Lett.\ B {\bf 653}, 424 (2007)
  [arXiv:0706.3873 [hep-ph]].

\bibitem{Barbieri:1987fn} 
  R.~Barbieri and G.~F.~Giudice,
  Nucl.\ Phys.\ B {\bf 306}, 63 (1988).

\bibitem{Birkedal:2004xi} 
  A.~Birkedal, Z.~Chacko and M.~K.~Gaillard,
  JHEP {\bf 0410}, 036 (2004)
  [hep-ph/0404197].

\bibitem{Bae:2012ir} 
  K.~J.~Bae, T.~H.~Jung and H.~D.~Kim,
  arXiv:1208.3748 [hep-ph].

\bibitem{Low:2009di} 
  I.~Low, R.~Rattazzi and A.~Vichi,
  JHEP {\bf 1004}, 126 (2010)
  [arXiv:0907.5413 [hep-ph]].

\bibitem{Carena:2012xa} 
  M.~Carena, I.~Low and C.~E.~M.~Wagner,
  JHEP {\bf 1208}, 060 (2012)
  [arXiv:1206.1082 [hep-ph]].


\end{thebibliography}
\end{document}